# Using Light to Polarize and Detect Electron Spins in Silicon

Xavier Marie[1,2*], Delphine Lagarde[1], Andrea Balocchi[1], Cédric Robert[1], Laurent Lombez[1], Pierre Renucci[1], Thierry Amand[1], Fabian Cadiz[3]

[1] *Université de Toulouse, INSA-CNRS-UPS, LPCNO, 135 Avenue Rangueil, 31077 Toulouse, France*
[2] *Institut Universitaire de France, 75231 Paris, France*
[3] *Laboratoire de Physique de la Matière Condensée, CNRS, Ecole Polytechnique, Institut Polytechnique de Paris, 91120 Palaiseau, France*
   * marie@insa-toulouse.fr

Despite decades of research, demonstration of all-optical detection and control of free electron spins in silicon remains elusive. Here, we directly probe the electron spin properties in bulk silicon by measuring the polarization of luminescence following circularly polarized light excitation. The experiments performed for both direct and indirect gap excitation allow not only an experimental determination of the optical selection rules in silicon for the different phonon-assisted transitions but they also lead to the measurement of the spin relaxation of electrons in conditions which are not accessible using transport techniques. We also measure the spin properties of free excitons in bulk silicon, a very little explored field.

Silicon is the reference semiconductor for the micro-nanoelectronics industry and more generally for our information society. Many prototype devices for processing and transmitting quantum information are based today on this strategic material [1,2]. It is also in silicon that Georges Lampel's pioneering experiments were carried out in 1968 on optical spin orientation in solids [3], wherein it was shown that polarized electronic spins are produced by optical pumping with circularly polarized light, yielding a dynamic polarization of $Si^{29}$ nuclei. The Lampel' studies have triggered very active research on the use of polarized light to study and control the spin of electrons in various III-V, II-VI, and group-IV semiconductors (eg, GaAs, CdTe, and Ge) and more recently in 2D semiconductors such as $MoS_2$ [4–9]. The optical techniques (luminescence, absorption, reflectivity) coupled with theoretical studies have made it possible to discover the optical selection rules and the spin relaxation mechanisms in a large variety of semiconductor structures [10,11]. Surprisingly these great advances did not concern free electrons in bulk silicon for which precise optical studies are more challenging since it is an indirect-gap semiconductor with weak phonon-assisted optical emission in the near infrared. Polarized luminescence, which is a direct probe of the free electron spin polarization, was never detected in bulk silicon following attempts to optically orient electron spins using polarized light [12]. The electron lifetime is expected to be much longer than the spin relaxation time ; the consequence of this unfavorable ratio is a predicted vanishing luminescence polarization measured under steady state conditions [13]. Despite theoretical and experimental studies carried out regularly for 50 years, the optical selection rules in silicon are still debated today since (i) no experimental data of polarized emission were available up to now and (ii) calculating the link between the polarized phonon-assisted luminescence and the electron spin polarization is complex as the radiative emission corresponds to transitions between electrons from the six conduction band valleys and heavy/light holes at the top of the valence band, Fig. 1(a) and 1(b) [14–18]. As it has not been possible so far to detect polarized luminescence after optical spin pumping, studies have focused on measurements made under strong external magnetic fields creating a steady state thermal spin polarization [14,19–21].

Spin relaxation times in bulk silicon could only be measured for localized electrons (or electrons trapped on defects) using pump-probe or electron spin resonance techniques and transport experiments requiring ferromagnetic material for the injection and electrical detection of free electron spins [22–28]. Elliot-Yafet spin relaxation mechanism due to the interaction of electron spin with phonons and defects has been identified as the dominant channel for spin-flip in these conditions [18,29]. Other processes based for instance on electron-hole exchange interaction (Bir-Aronov-Pikus) which plays a key role for other semiconductors were never evidenced in silicon [4]. Exciton spin properties in silicon have also never been measured either in the absence of magnetic field since the spin polarization of exciton was not evidenced following circularly polarized excitation light.

We were able in this work to detect circularly polarized silicon luminescence after optical spin orientation under both direct and indirect gap excitation conditions. This was possible thanks to ultra-stable low-temperature optical spectroscopy set-ups allowing the measurement of polarization degrees as low as $10^{-4}$. These experiments lead to (i) the determination of the optical selection rules for the different transitions assisted by phonon emission (TO, LO and TA) and (ii) the measurement of the spin relaxation time of free electrons.

We present the results on two typical samples. The first sample Si(p) is a p-type silicon wafer with a doping density of about $10^{18}$ cm$^{-3}$. The second sample Si(i) is non-intentionally doped with a typical residual doping of $10^{13}$ cm$^{-3}$ (see Supplemental Material, SM) [30]. Unless stated, the measurements have been performed with samples in optical cryostats at low temperature (T=10 K).

*Direct gap excitation*

First we investigate the polarized photoluminescence (PL) of the p-type silicon (phonon-assisted indirect gap recombination) following circularly-polarized laser excitation resonant with the direct gap transition. The advantage of the direct gap excitation is that it does not involve any phonon and large initial spin polarization of photogenerated electrons are expected [17,31]. The excitation/detection scheme is shown in Figure 1(b). The sample is excited by 1.5 ps pulses at a wavelength of 375.4 nm (3.302 eV). This excitation wavelength maximizes the spin polarization of Γ electrons as shown in photo-emission spectroscopy measurements [31] . The details on the optical spectroscopy set-up are given in the SM.

Figure 1(c) displays the right ($I^+$) and left ($I^-$) circularly polarized luminescence spectra and the corresponding luminescence polarization $P_c = (I^+ - I^-)/(I^+ + I^-)$. The indirect gap luminescence is dominated by the TO-phonon assisted recombination (1.081 eV) and a lower intensity TA-phonon assisted component (1.120 eV), see Supplemental Material S1 for more information on the identification of the PL lines. Let us recall that TO, LO and TA phonon energies are 58, 56 and 18.5 meV respectively [32]. Remarkably the TO and TA luminescence components exhibit a polarization $P_c^{TO} = 0.91 \pm 0.02$ % and $P_c^{TA} = 1.17 \pm 0.02$ % respectively. We also observe a clear negative polarization of about - 0.3% on the high energy tail of the TO luminescence peak. Considering the large p-type doping of the sample, the luminescence polarization observed in Fig. 1(c) can be unambiguously assigned to the optical recombination of spin-polarized free electrons with unpolarized holes [32].

The measurement of a significant luminescence polarization of the order of ~ 1% is a consequence of the very specific excitation conditions. The measured luminescence circular polarization writes simply [10] :

$$P_c^{TO/TA} = \frac{P_0 \cdot \phi^{TO/TA}}{1 + \frac{\tau}{\tau_s}} \qquad (1)$$

, where $P_0$ is the X valley electron spin polarization following photogeneration and relaxation, $\phi^{TO/TA}$ is a factor reflecting the spin selection rules associated to TO or TA phonon-assisted emission, τ and τ$_s$ are respectively the electron lifetime and spin relaxation time.

The different features observed in Fig. 1(c) contain essential information on the optical selection rules in silicon. The parameter $\phi^{TO/TA}$ (<1) in equation (1) is a consequence of the symmetry of the electron wave functions and TO/TA phonons involved in the optical transitions [15,16]. As $P_0$, τ and τ$_s$ are identical for both phonon-assisted transitions, the measured values of the luminescence polarization $P_c^{TO}$ and $P_c^{TA}$ allow us to determine directly the ratio between $\phi^{TA}$ and $\phi^{TO}$. We find $\phi^{TA}/\phi^{TO} = 1.28$. Our measurements are in very good agreement with calculations performed by Li and Dery who predicted $\phi^{TO} = 0.188$ and $\phi^{TA} = 0.235$, corresponding to a ratio $\phi^{TA}/\phi^{TO} = 1.25$ [16] . Note that previous theoretical calculations yield larger values with $\phi^{TO} = 0.4$ [15]. We also measure a ratio between the TO and TA luminescence intensities $I^{TO}/I^{TA} \sim 13$. Remarkably, this experimental ratio is also in excellent agreement with the Li and Dery calculations that predicted a ratio between the TO and TA luminescence intensity of 12.8. Finally, the negative polarization observed in Fig. 1(c) close to the TO line is attributed to a weak contribution of the LO assisted radiative recombination of spin polarized electrons. This LO-assisted luminescence lies about 2 meV above the TO line. As it is ~ 7 times weaker than the TO line [34], no associated peak is observed in luminescence due to the broadening of the strong TO line but is has a clear effect on the measured circular polarization. As a direct consequence of the optical selection rules in silicon, this LO-assisted luminescence contribution yields an opposite sign of the luminescence polarization. In contrast the TA and TO phonons share the same symmetry and therefore their associated optical transitions have the same polarization sign [16]. The Supplemental Material S2 shows a detailed comparison of the measured and calculated luminescence polarization which confirm these selection rules for TO, LO and TA transitions.

We emphasize that a significant polarization (~1%) can be observed in Fig. 1(c) thanks to the very specific excitation conditions used here. The high laser repetition rate (80 MHz) pulsed excitation yields a very high photogenerated carrier density due to carrier accumulation, as already observed in silicon [35]. This leads to an unusually very short electron lifetime $\tau$ – in the nanosecond range - governed by Auger effect [36]. Equation (1) shows that the consequence is a large increase of the luminescence polarization, compared to the situation of the usual silicon electron lifetime lying in the microsecond range [37]. As expected, if we use similar excitation energy and average power from a cw excitation laser (instead of a high repetition rate pulsed laser) we record a much smaller polarization $P_c$ as a consequence of a much longer electron recombination time (Supplemental Material S3); the Auger effect is negligible due to the much smaller photogenerated carrier density in this cw excitation regime.

A decisive experiment to further prove that the circular polarization of the luminescence is linked to the electron spin polarization consists in measuring the drop in the luminescence polarization due to the application of a transverse magnetic field (Hanle effect) [4]. Indeed, the magnetic field B applied perpendicular to the electron spin will induce spin precession with a frequency $\omega = g\mu_B B/\hbar$ ; here, g is the electron Landé g factor, $\mu_B$ is the Bohr magneton, and $\hbar$ the Planck constant. If the precession period is of the order, or smaller than the electron spin lifetime, the average value of the electron spin polarization $P_s$ will decrease, following the simple Lorentzian dependence: $P_s(B) = \frac{P_s(0)}{1+(\omega T_s)^2}$, with the electron spin lifetime $T_s$ defined by $1/T_s=1/\tau+1/\tau_s$. As the electron g factor in silicon is well known (g $\simeq$ 2 ) [38], this technique allows us to determine the electron spin relaxation time $\tau_s$ if the electron lifetime $\tau$ is measured separately [4].

In order to avoid strong Auger and non-linear effects induced by the very large photogenerated carrier density in the high frequency pulsed excitation conditions used in Fig. 1, we performed the Hanle experiment using a cw excitation laser resonant with the direct gap. Although this excitation condition avoids collective effects, it yields a much weaker luminescence polarization (typically 100 times weaker than the one measured in Fig. 1) as a consequence of a much longer electron lifetime. Luminescence polarization measurement then becomes a challenge; despite high experimental uncertainties, it is nevertheless possible to measure the dependence of the polarization as a function of the magnetic field thanks to modulation techniques (see SM). Figure 2a displays the measured Hanle curve. The full line is a fit yielding an electron spin lifetime $T_s$=0.240 $\pm$ 0.03 ns. The electron lifetime $\tau$ has been measured separately using time-resolved photoluminescence, as shown in Fig. 2(b). From the luminescence decay time, we can determine $\tau$= 210 $\pm$10 ns. Note that for the low pulsed laser excitation rate used here (400 kHz), the luminescence intensity depends linearly with the excitation power; thus Auger effects play a negligible role (in a similar way as for the cw excitation conditions of the Hanle measurement).

These experiments therefore allow us to directly measure the spin relaxation time of electrons in silicon by an all-optical technique : we find $\tau_s \sim 0.240$ ns. We obtain a much shorter time than those obtained by transport or electron spin resonance techniques, for which values in the range of few hundreds of ns have been determined in n-doped silicon [27,39] . While for these measurements the dominant spin relaxation mechanism is linked to the interaction with phonons (Elliot-Yafet type), our measurements carried out in a p-doped sample (~ $10^{18}$ cm$^{-3}$) highlight the importance of the electron spin relaxation mechanism induced by exchange interaction with holes (Bir-Aronov-Pikus type), never observed for silicon before [40–43]. We also observe the effect of the transverse magnetic field on the luminescence polarization in the very high excitation density regime of Figure 1 (Supplemental Material S4).

*Indirect gap excitation*

The results presented so far correspond to photogeneration of spin-polarized electrons in the Γ valley (direct gap excitation) and the detection of luminescence polarization linked to indirect gap phonon-assisted recombination. We show now that it is also possible to photo-generate directly electron spins close to the X-valley using indirect gap excitation and detecting the corresponding circular polarization emission of the indirect gap emission (Fig. 3(a)). The luminescence process is similar to the one already described in Fig. 1 and 2. However there are two key differences for the photogeneration :

(i) the optical absorption coefficient α drops by 5 orders of magnitude, passing from a direct gap excitation to an indirect gap excitation assisted by TA phonon emission (α~ 10 cm$^{-1}$) for the onset of absorption [44] ; being at low temperature, the processes involving phonon absorption can be neglected . Thus, the luminescence intensity will be much weaker.
(ii) the phonon-assisted photogeneration of electrons has to be considered for the determination of the optical selection rules which dictate the maximum electron spin polarization in the X valley [18,45] . The selection rules related to the electron-phonon coupling therefore intervene twice (Fig. 3(a)), leading to very weak luminescence polarization values.

Figure 3(b) displays the dependence of the TO-assisted PL intensity and corresponding circular polarization as a function of the excitation laser energy (indirect gap excitation). The cw excitation laser is circularly polarized (T=4 K). As expected the onset of absorption corresponds to the energy gap $E_g$ plus the energy of the TA phonon ($E_g$+TA) (see Supplemental Material S1 for the determination of $E_g$) [33]. For larger excitation energy, the luminescence intensity rises strongly, in particular when the TO phonon-assisted absorption occurs ($E_g$+TO). When the spin-polarized electrons are photogenerated at the bottom of the X valley ($E_{laser}=E_g+TA$), the measured luminescence polarization is $P_c \sim 6.\,10^{-4}$ and then drops drastically when the laser energy increases. Here, the luminescence polarization linked to the recombination of spin-polarized electrons is extremely low. To eliminate all polarization artefacts which are of the same order of magnitude as spin polarization, the polarization plotted in Fig. 3(b) corresponds to the difference between a measurement at B=0 and a measurement at B=50 mT (for this value of transverse magnetic field the average spin polarization vanishes, see Fig. 2(a)). In Fig. 3(c), the excitation energy dependence of the measured luminescence polarization is compared to the calculated photogenerated electron spin polarization $P_{gene.}$ in the X valley following indirect gap optical spin injection [18] ; these calculations are based on an empirical pseudopotential model and an adiabatic bond charge model for electron and phonon states respectively. We clearly see that the drop of the measured luminescence polarization can be well explained by the strong decrease of the initial photogenerated spin polarization. For excitation energies larger than ~30 meV above the gap, the phonon-assisted optical spin injection process vanishes. These results demonstrate that very resonant excitation conditions are required in order to photogenerate significant electron spin polarization in bulk silicon (indirect gap excitation), as a consequence of the small spin-orbit splitting and spin mixing effects in the valence bands. Note that the polarization drop in Fig. 3(c) occurs for excitation energies lower than those capable of populating the spin-orbit split off valence band.

By performing similar experiments in a non-intentionally doped sample with low residual doping (~$10^{13}$ cm$^{-3}$), we show that that the optical orientation technique is also a very powerful tool to investigate the free exciton spin physics in bulk silicon, a very little explored field [14]. We have measured the dependence of the TO-assisted free exciton PL intensity and corresponding circular polarization as a function of the excitation laser energy (Supplemental Material S5). Free exciton spin polarization following polarized light excitation is clearly evidenced. For comparison, we also measured the polarized luminescence obtained in strong external longitudinal magnetic field (a few Teslas in Faraday configuration) as a consequence of thermal electron spin polarization. In that case the imbalance between the spin up and spin down photogenerated electron populations result from the Zeeman splitting induced by the external magnetic field. These experiments confirm the optical selection rules evidenced above, in particular the opposite sign of the circular polarization for TO and LO-assisted optical recombination (see SM S6).

In summary we have demonstrated that circularly polarized light photogenerates in bulk silicon spin-polarized electrons which can be directly probed by luminescence spectroscopy. These experiments performed for both direct and indirect gap excitation conditions allow us to evidence the optical selection rules for the different phonon-assisted optical transitions and measure the free electron or exciton spin properties in conditions inaccessible to transport experiments. These precise optical measurements of the spin properties of electrons should also make it possible to evidence the suppression of spin relaxation in strained silicon [46–48].

**Acknowledgments :**
We thank Georges Lampel for very stimulating discussions. Part of this work has been supported by the France 2030 government investment plan managed by the French National Research Agency under grant reference PEPR SPIN – SPINMAT ANR- 22-EXSP-0007.

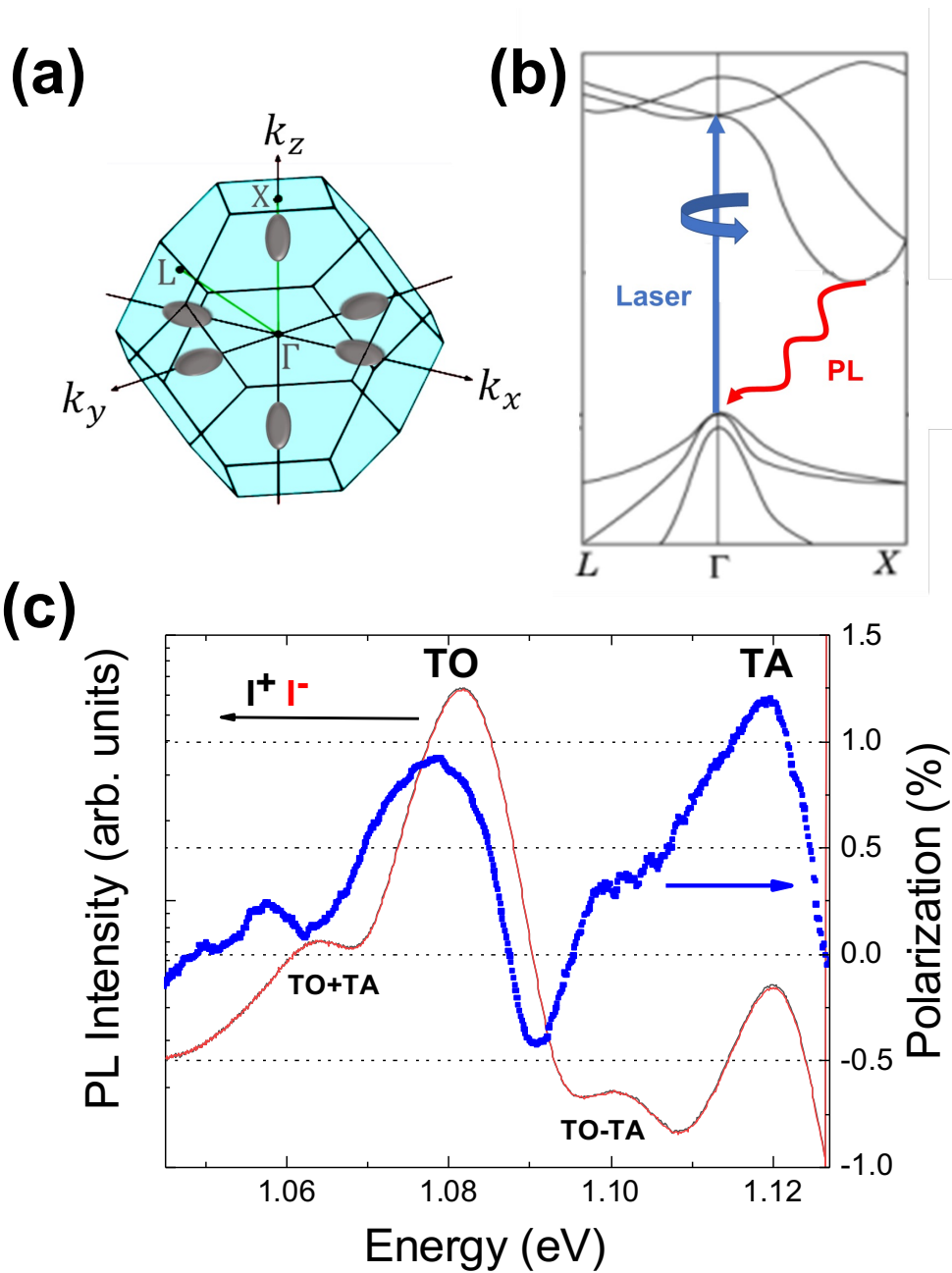

**Fig.1 Optical spin orientation in Silicon following <u>direct gap</u> pulsed laser excitation – p-doped Silicon – (a)** Schematics of Silicon first Brillouin zone with the six conduction band valleys located close to the X symmetry point ; **(b)** Simplified band structure displaying the laser excitation set to the direct gap excitation (3.302 eV, blue arrow) and the indirect gap phonon-assisted luminescence (red arrow); **(c)** Right ($I^+$) and Left ($I^-$) circularly-polarized luminescence intensity (log. scale) and the corresponding circular polarization $P_c=(I^+-I^-)/(I^++I^-)$, the pulsed excitation laser is circularly-polarized (pulse width ~1.5 ps, 80 MHz repetition rate). Significant polarizations are observed for both TO and TA phonon-assisted lines corresponding to recombination of photogenerated spin-polarized electrons with unpolarized holes (the weaker intensity TO+TA and TO-TA peaks result from two-phonons assisted recombination processes).

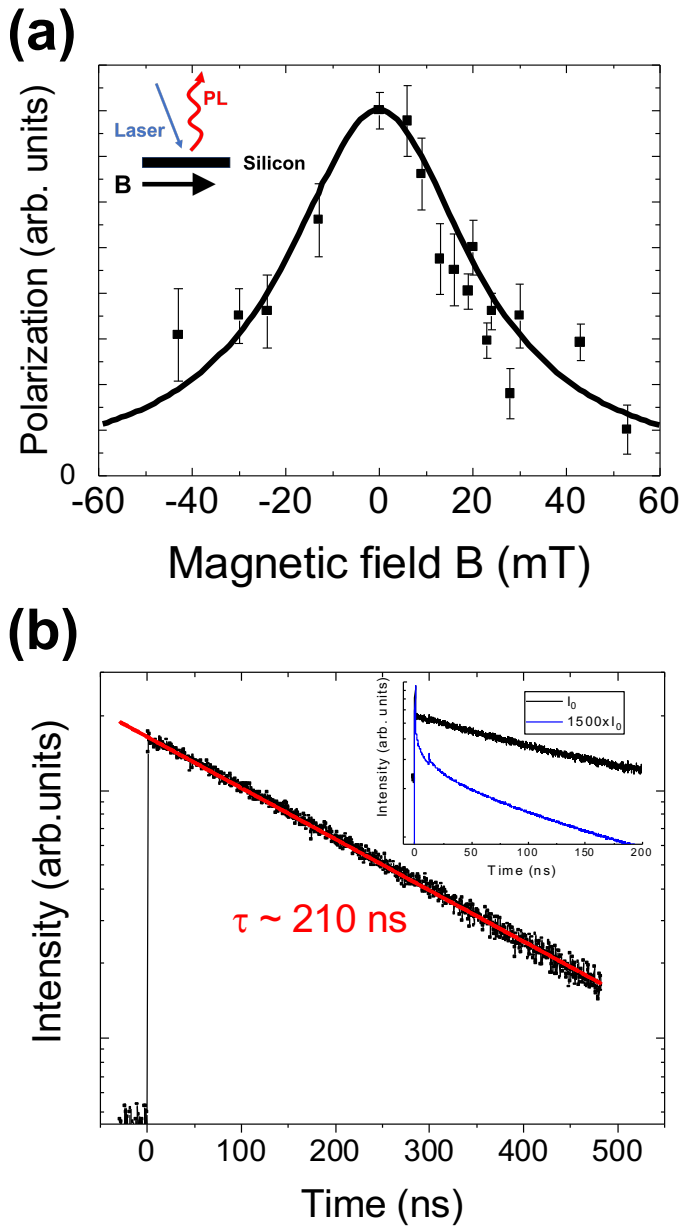

**Fig.2 Photogenerated electron spin dynamics for <u>direct gap</u> laser excitation – p-doped Silicon –** **(a)** Hanle curve : Normalized luminescence polarization of the TO luminescence as a function of a transverse magnetic field B (B lies in the sample plane and perpendicular to the light propagation axis, Voigt configuration, see the inset); the full line is a Lorentzian fit yielding an electron spin relaxation time of $\tau_s$=0.24 ns. The cw laser excitation intensity is about 0.5 kW.cm$^{-2}$. **(b)** T=5 K. Time-resolved photoluminescence intensity, the red line is an exponential fit giving an electron lifetime $\tau$=210 ns (the pulsed laser energy is 3.262 eV, the repetition rate 400 kHz, the average intensity I~7 W.cm$^{-2}$). Inset : Luminescence kinetics recorded for larger repetition rate (4 MHz) and two laser excitation intensities : $I_0$~ 2 W.cm$^{-2}$ and 1500x$I_0$ . For the low excitation intensity $I_0$, we measure again a decay time of ~200 ns, while for higher excitation (1500x$I_0$), the drop of the electron lifetime due to Auger effect is clearly evidenced (excitation conditions close to the ones used in Fig. 1(c)).

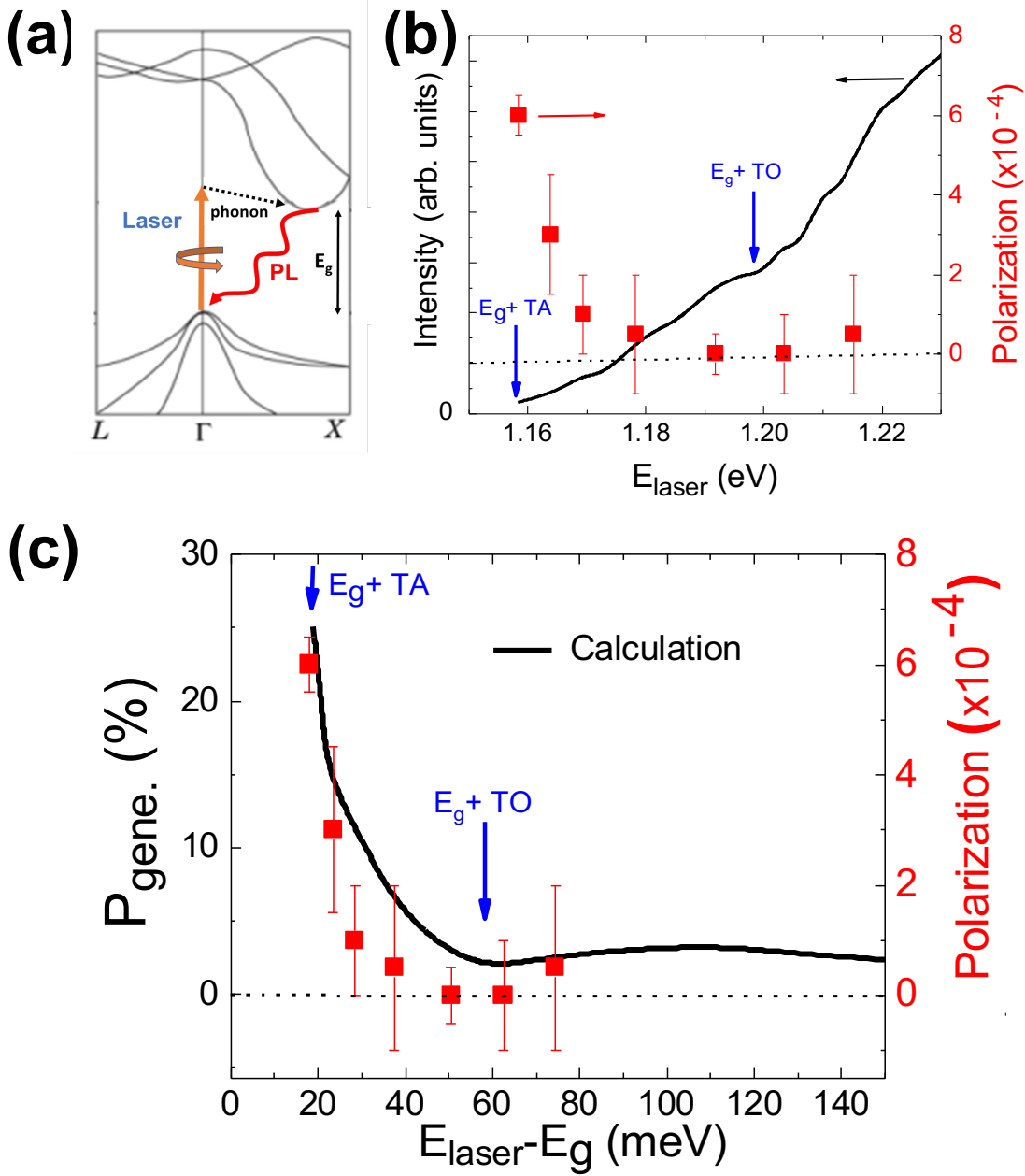

**Fig.3 Optical spin orientation in Silicon following <u>indirect gap</u> laser excitation – p-doped Silicon – (a)** Silicon band structure displaying the phonon-assisted laser excitation (orange arrow) and the indirect gap phonon-assisted luminescence (red arrow). **(b)** Polarized Excitation of Photoluminescence : luminescence polarization of the TO line as a function of the cw excitation laser energy. The full line displays the variation of the luminescence intensity (the blue vertical arrows indicate the onset of TO and TA assisted phonon absorption). **(c)** Comparison of the variation of the measured polarization and the calculated photogenerated electron spin polarization $P_{gene.}$ [18] as a function of the photogenerated carrier excess energy with respect to the indirect gap $E_g=1,139$ eV.

**Using Light to Polarize and Detect Electron Spins in Silicon**

Xavier Marie[1,2*], Delphine Lagarde[1], Andrea Balocchi[1], Cédric Robert[1], Laurent Lombez[1], Pierre Renucci[1], Thierry Amand[1], Fabian Cadiz[3]

[1] *Université de Toulouse, INSA-CNRS-UPS, LPCNO, 135 Avenue Rangueil, 31077 Toulouse, France*
[2] *Institut Universitaire de France, 75231 Paris, France*
[3] *Laboratoire de Physique de la Matière Condensée, CNRS, Ecole Polytechnique, Institut Polytechnique de Paris, 91120 Palaiseau, France*
  * marie@insa-toulouse.fr

**Contents:**

## Section 1. Samples and experimental set-ups

<u>Bulk silicon samples</u>

We present the results obtained in two samples. The first sample Si(p) is a p-type Boron doped (001) silicon wafer - 250 micrometer thick - from BT electronics (surface finish mirror polished); resistivity 0.03-0.05 Ohm.cm corresponding to a doping density of about $10^{18}$ cm$^{-3}$. The second sample Si(i) is a non-intentionally doped silicon wafer with a measured resistivity at room temperature of about 1600 Ohm.cm corresponding to a typical residual doping of $10^{13}$ cm$^{-3}$.

<u>Optical measurements</u>

All the measurements were performed at low temperature using closed-cycled Helium cryostats with temperatures in the range 4-10 K. In order to optimize the accuracy of the polarization measurements, different set-ups were used depending on the excitation conditions.

*(a) Direct gap and pulsed laser excitation (Figure 1 of the main text):*
The excitation laser wavelength was fixed at λ=375.4 nm (3.302 eV). For this excitation energy, a spin polarization of Γ electrons of about 10% was measured by photo-emission spectroscopy in a silicon sample with the same p-type doping ($10^{18}$ cm$^{-3}$) as the one investigated here [1]. The results presented in Fig. 1(c) were obtained using a frequency doubled mode-locked Ti-sapphire laser (Tsunami from Spectra Physics) with a 1.5 ps pulse width and a repetition rate of 80 MHz. The average excitation power intensity on the sample was typically 2 kW.cm$^{-2}$. The circular polarization of the exciting beam was set by a combination of a linear polarizer and a quarter-wave plate, while the degree of the luminescence circular polarization was obtained by passing the PL signal through a rotating quarter-wave plate followed by a linear polarizer. The PL signals was recorded using a liquid-nitrogen-cooled InGaAs array detector (Pylon-IR, Princeton Instruments) coupled to a spectrometer (SpectraPro HRS -500). The typical acquisition time of the PL spectra for each polarization was a few tens of minutes.

For the time-resolved photoluminescence measurements (T=5 K), a pulse picker was used to reduce the repetition rate of the frequency-doubled Ti:Sa laser from 80 MHz to 4 MHz or 400 kHz (see Fig. 2). The near-infrared TRPL was dispersed by a spectrometer (SpectraPro HRS 500mm from Princeton Instruments) which exit is coupled to a single-mode optical fiber for the measurement of the PL kinetics. This fiber was connected to a superconducting nanowire single photon detector (Single Quantum) in combination with a time- correlated single photon counting system (PicoHarp from PicoQuant). The temporal shape of the detected luminescence signal was acquired as a histogram made of 65000 bins with a bin width of 512 ps. The Instrument-Response Function (IRF) width of the overall setup is in the order of 1 ns. We found similar luminescence decay times for laser excitation energies above or below the direct gap. This proves that surface recombination plays minor roles in both cases.

*(b) Direct gap excitation and cw laser excitation (Hanle measurements):*
For the Hanle measurements (Figure 2 of the main text), the sample was excited by a stabilized cw diode laser (L375P70MLD, ThorLabs ) , with very similar wavelength (376 nm) as the one used for pulsed excitation conditions (Figure 1). The excitation power is ~ 30 mW and the laser spot diameter about 75 μm. The sample temperature is T=10 K. A transverse magnetic field is applied in the plane of the sample (Voigt configuration) using two permanent Ferrite magnets and the field strength on the sample surface had been varied by changing the magnet's distance from the sample. The excitation laser polarization is modulated at a frequency of f=73 Hz using a rotating λ/4 plate. The PL signal after passing a λ/4 plate, a polarizer and a band-pass filter centered on the TO emission line is detected by an InGaAs avalanche photodetector (APD410C/M, ThorLabs) coupled to a lock-in amplifier (SR830, Stanford Research Systems). In this configuration, the polarization state after the rotating wave-plate on the excitation path will periodically change from linear to circular right, linear, circular left and back to linear, twice per rotation. By demodulating the photodetector signal at a frequency 2*f* using the lock-in amplifier, the resulting signal is proportional to ($I^+ - I^-$), *i.e.* the PL circular polarization [2]. An additional mechanical chopper coupled to a second lock-in amplifier yields a signal that is proportional to ($I^+ + I^-$) signal. This technique, together with the use of a stabilized diode laser and stabilized sample temperature ($\pm 0.1\ K$) allows us to measure a signal proportional to $P_c=(I^+-I^-)/(I^++I^-)$ for very small PL circular polarization $P_c$ down to a few $10^{-4}$.

*(c) Indirect gap cw excitation conditions :*
A stabilized tunable external cavity diode laser (Sacher, TEC-500-1060-030) was used for the indirect gap phonon-assisted excitation experiments with laser wavelengths in the range 990-1075 nm (measurements of Fig. 3 of the main text and Fig. S5 of the Supplemental Information). The polarization and detection techniques were similar as the time-integrated measurements in conditions (a) presented above. Due to the small difference between the laser excitation and detection energies, particular attention is paid to polarized Raman lines which have nothing to do with the optical spin orientation of the electrons [1,3]

*(d) Indirect gap cw excitation conditions – longitudinal magnetic field (Faraday) :*
The measurements of Fig. S6 were obtained using a magneto-PL set-up at T = 5 K and magnetic fields perpendicular to the sample surface (Faraday geometry) up to 6 T (closed -cycle Helium Magneto-cryostat MycryoFirm) . The sample was excited by a He-Ne laser (1.96 eV) with linear polarization and both circular σ+ and σ- polarized PL signals were detected using a λ/4 plate and an analyzer in the detection path. The average laser intensity is about 5 kW.cm$^{-2}$ (The excitation/detection spot diameter is about 1 μm). The PL emission is dispersed by a monochromator (Princeton Instruments, Acton SpectraPro SP-2500) and detected by a liquid-nitrogen-cooled InGaAs array detector.

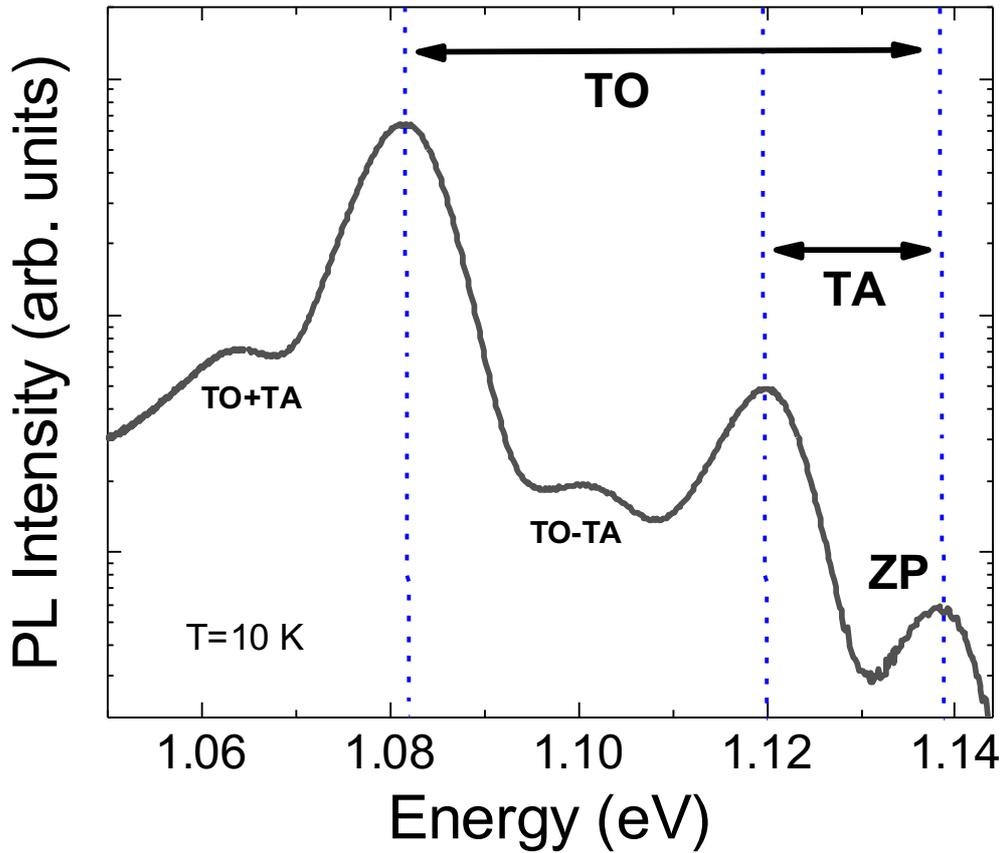

**Fig. S1. Photoluminescence spectrum following <u>direct gap</u> laser excitation – p-doped Silicon –** T=10 K. The excitation laser energy is 3.303 eV. The two main peaks energies (1.081 and 1.120 eV) are in agreement with previous measurements performed in heavily p-doped samples [6] ; it corresponds to the TO and TA phonon-assisted optical recombination between free conduction band electrons and holes in the acceptor impurity band [7] ; the horizontal arrows correspond to the energy of the TO (58 meV) and the TA (18,5 meV) phonons [5]. The shape of the spectrum is unchanged on a large excitation density range, confirming the band to band nature of the recombination. The scattering on impurities yield a weak zero-phonon optical recombination (ZP) observed at 1.139 eV which allow us to determine the indirect optical gap energy $E_g$ [8].

**Section 2. Optical selection rules : comparison Experiments and Theory**

In order to compare the measured and theoretical circular polarization of the luminescence, we have considered the spin-dependent phonon-assisted optical transitions calculated by Li and Dery for TO, TA and LO phonons [4]. Table 1 displays the calculated relative intensities and corresponding circular polarization for the three transitions assuming a 100% electron spin polarization in the X valley.

The photoluminescence spectrum obtained after direct gap excitation (Figure 1 of the main text) is simply modelled considering three Gaussian functions centered on the three main phonon-assisted transitions with the well-known TO, LO and TA phonon energies 58, 56 and 18.5 meV [5]. The fitting procedure is the following: first we fit the PL intensity and the circular polarization of the TA-assisted transition (1.120 eV). The fitting parameter for the circular polarization takes into account the loss of electron spin polarization induced by the energy relaxation from the $\Gamma$ to X valley. Then we use the theoretical intensity ratios and circular polarization displayed in Table 1 to compute the TO and LO luminescence components $I^+$, $I^-$ and $I=I^+ + I^-$. The black, orange and pink dotted lines in figure S2 are the corresponding calculated TO, LO and TA PL intensity components I (the Gaussian broadening parameters are 7,5, 12 and 7,5 meV respectively). The bold line is the sum of the three contributions. We observe a good agreement between the experimental and calculated PL intensities. As nothing is known from a theoretical point of view about the spin-dependent two-phonons optical transitions, we have not considered the TO+TA and TO-TA lines and the results of the calculation in these two spectral regions have to be discarded.

The dotted blue line is the corresponding calculated PL polarization. We emphasize that no additional fit parameter was included. The calculated polarization reflects well the measured one at the peak of the main TO and TA lines and it also reproduces the negative polarization observed on the high energy tail of the main TO line (~1,090 eV) due to the contribution of the LO-assisted phonon transition with opposite polarization.

|  | TA | TO | LO |
|---|---|---|---|
| **PL Intensity** | 1 | 12.86 | 1.73 |
| **Polarization (%)** | 23.5 | 18.8 | -27.7 |

**Table 1**. Calculated relative light intensities and circular polarization of TA, TO and LO phonon-assisted optical transitions of spin-down electrons at the bottom of the conduction band (X valley) to heavy and light hole states at the top of the valence band ($\Gamma$ valley). The calculations are the sum of contributions from all six CB valleys. The light intensities are normalized with respect to the TA intensity [4].

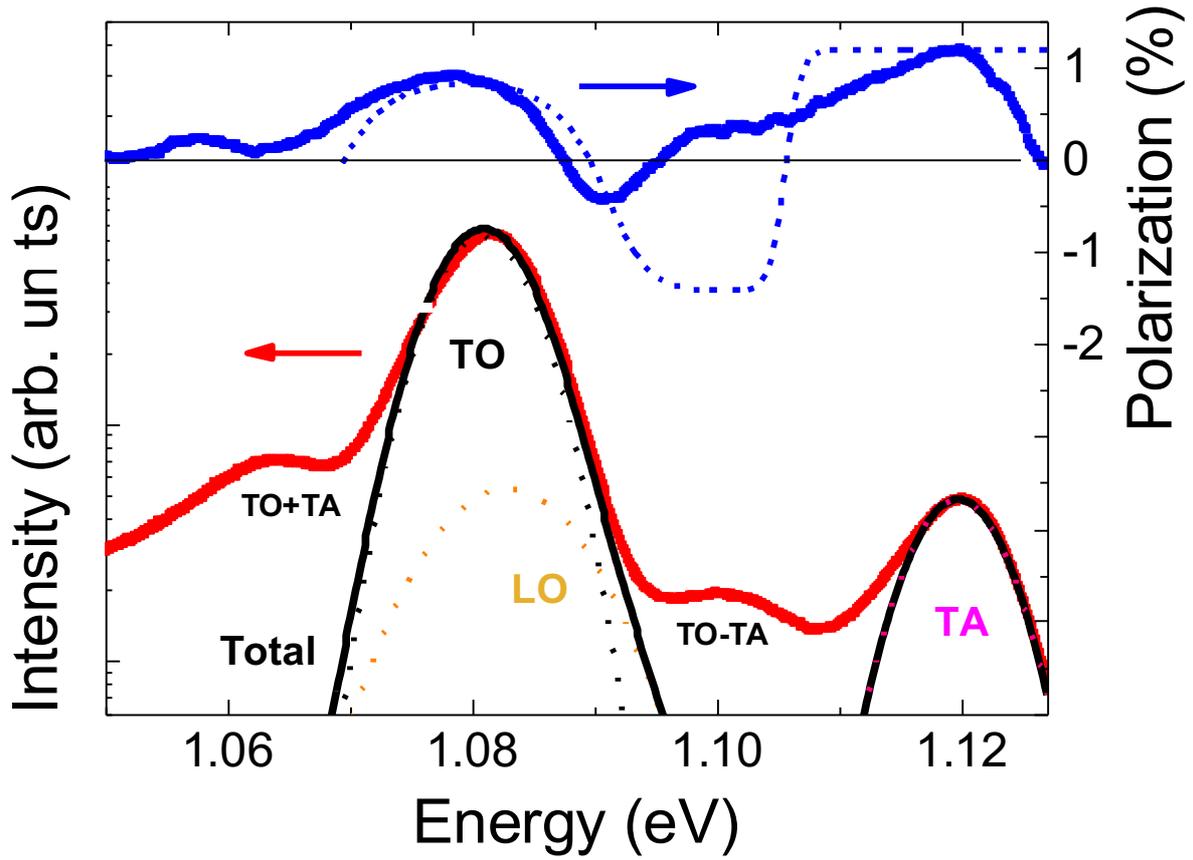

**Fig. S2. Optical selection rules : comparison between Experiments and Theory – p-doped Silicon –** Pulsed excitation (1.5ps). Repetition rate: 80MHz. Measured photoluminescence intensity (I=I$^+$+I$^-$, red) and the corresponding circular polarization P$_c$=(I$^+$-I$^-$)/(I$^+$+I$^-$), blue, as in Fig.1c of the main text. The excitation laser is circularly-polarized. The black, orange and pink dotted lines are the calculated TO, LO and TA PL intensity components I (the bold line is the sum of the three contributions.) and the dotted blue line is the calculated PL polarization following spin-dependent phonon-assisted optical selection rules [4], see section 2 for the details of the calculation. We emphasize that the calculation does not include two-phonon assisted optical recombination processes; so the comparison between experiments and calculations in the TO+TA and TO-TA energy range is meaningless.

**Section 3. Optical spin orientation following direct gap laser excitation – p-doped Silicon ( cw excitation)**

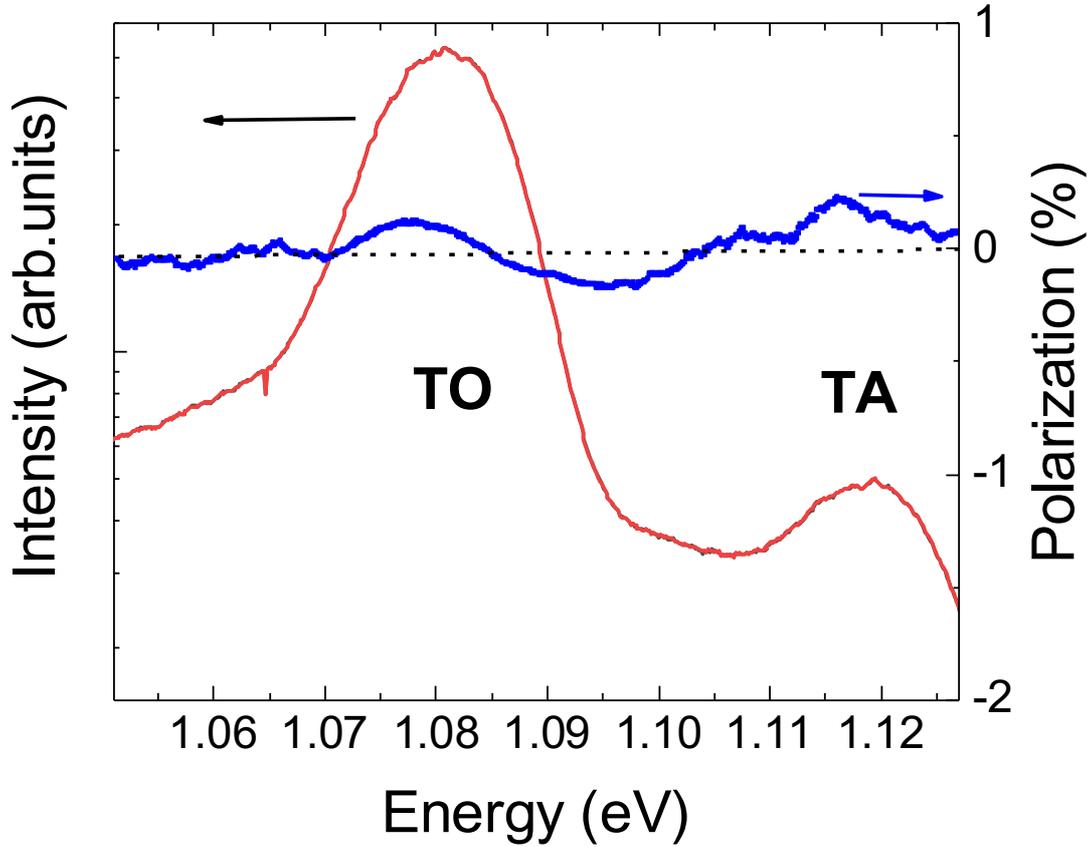

**Fig. S3. Optical spin orientation in Silicon following direct gap laser excitation – p-doped Silicon** – cw excitation. Right ($I^+$) and Left ($I^-$) circularly-polarized luminescence intensity and the corresponding circular polarization $P_c=(I^+-I^-)/(I^++I^-)$, the excitation laser is circularly-polarized and its energy is 3.293 eV.

In contrast to the measurements in Figure 1c where high repetition rate (80 MHz) pulsed laser (1.5 ps) was used, the excitation laser operates here in cw mode with similar average intensity. This yields a much smaller photogenerated carrier density and as a consequence a longer electron lifetime due to the suppression of the Auger effect [9] . These excitation conditions lead to much smaller measured polarization. With the set-up used here with no modulation techniques, the measured polarization (~0.1%) is within the experimental uncertainty.

**Section 4. Effect of a transverse magnetic field on the photoluminescence polarization ; direct gap pulsed laser excitation – p-doped Silicon**

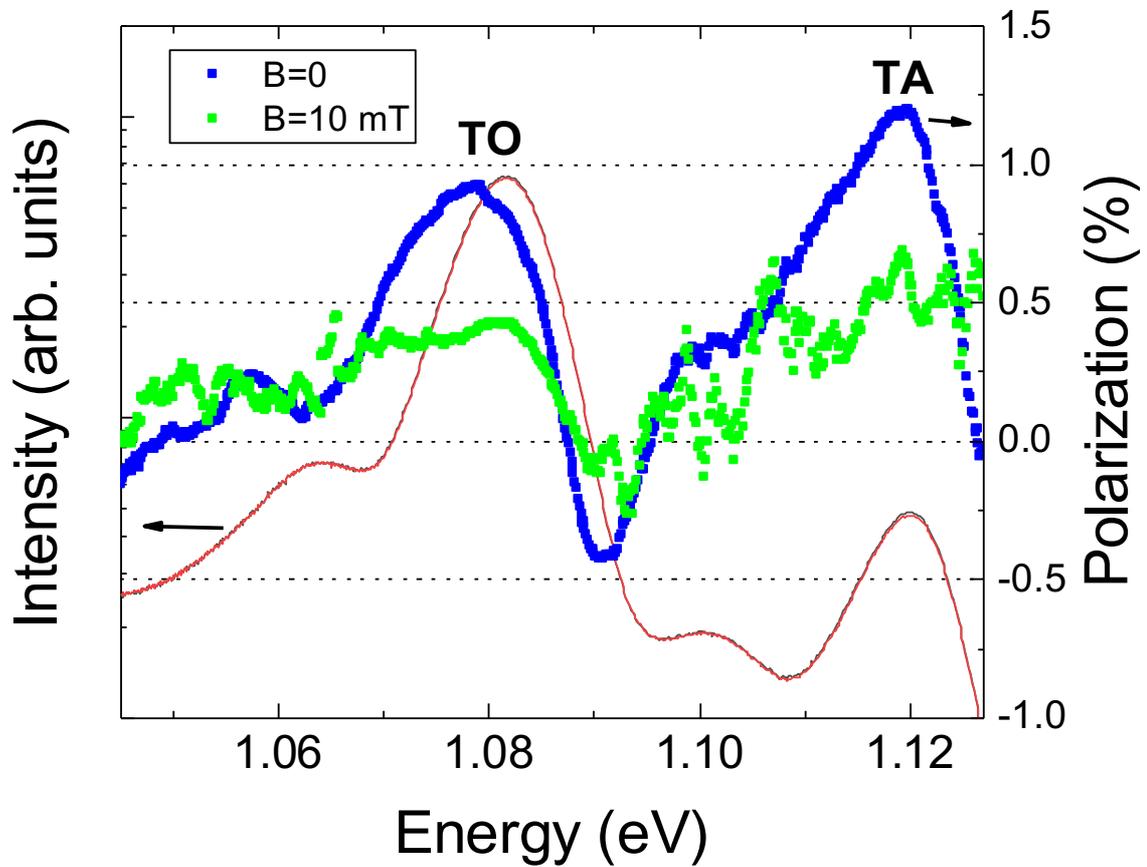

**Fig. S4. Effect of a transverse magnetic field on the photoluminescence polarization ;** direct gap pulsed laser excitation – p-doped Silicon – The black and red lines are the right ($I^+$) and left ($I^-$) circularly-polarized luminescence intensity (B=0 T); the blue and green dots correspond to the measured circular polarization $P_c=(I^+-I^-)/(I^++I^-)$, for B=0 and B=10 mT respectively. The magnetic field B lies in the sample plane and perpendicular to the light propagation axis, Voigt configuration. The excitation/detection conditions are identical to the ones of Fig. 1c in the main text (pulsed laser).

## Section 5. Optical spin orientation of exciton, indirect gap laser excitation – Intrinsic Silicon

We have investigated a non-intentionally doped sample with low residual doping ($\sim 10^{13}$ cm$^{-3}$). Figure S5(a) displays the cw photoluminescence spectrum at T=5 K .The luminescence is dominated by the free exciton (FX) luminescence assisted by TO phonon emission ($E_{FX-TO}$= 1.097 eV) [10,11] ; the weaker TA-phonon assisted FX component is also clearly observed at $E_{FX-TA}$= 1.136 eV. Let us recall that the free exciton binding energy in bulk silicon is 14 meV [12] . For larger excitation powers, a new line appears ~ 15 meV below FX corresponding to the electron-hole droplet recombination [13] . Figure S5(b) presents the dependence of the TO-assisted free exciton PL intensity and corresponding circular polarization as a function of the excitation laser energy. A polarization of about ~$3.10^{-4}$ can be observed for $E_{laser}$=1.186 eV ; it vanishes for larger excitation energy. Surprisingly no polarization can be detected within our experimental uncertainties when the free exciton is photogenerated resonantly following TA phonon assisted absorption ($E_{laser}$=1.175 eV). This could be the signature of a very efficient exciton spin relaxation mechanism which occurs only for cold free exciton in silicon as it was observed for other semiconductor structures [14].

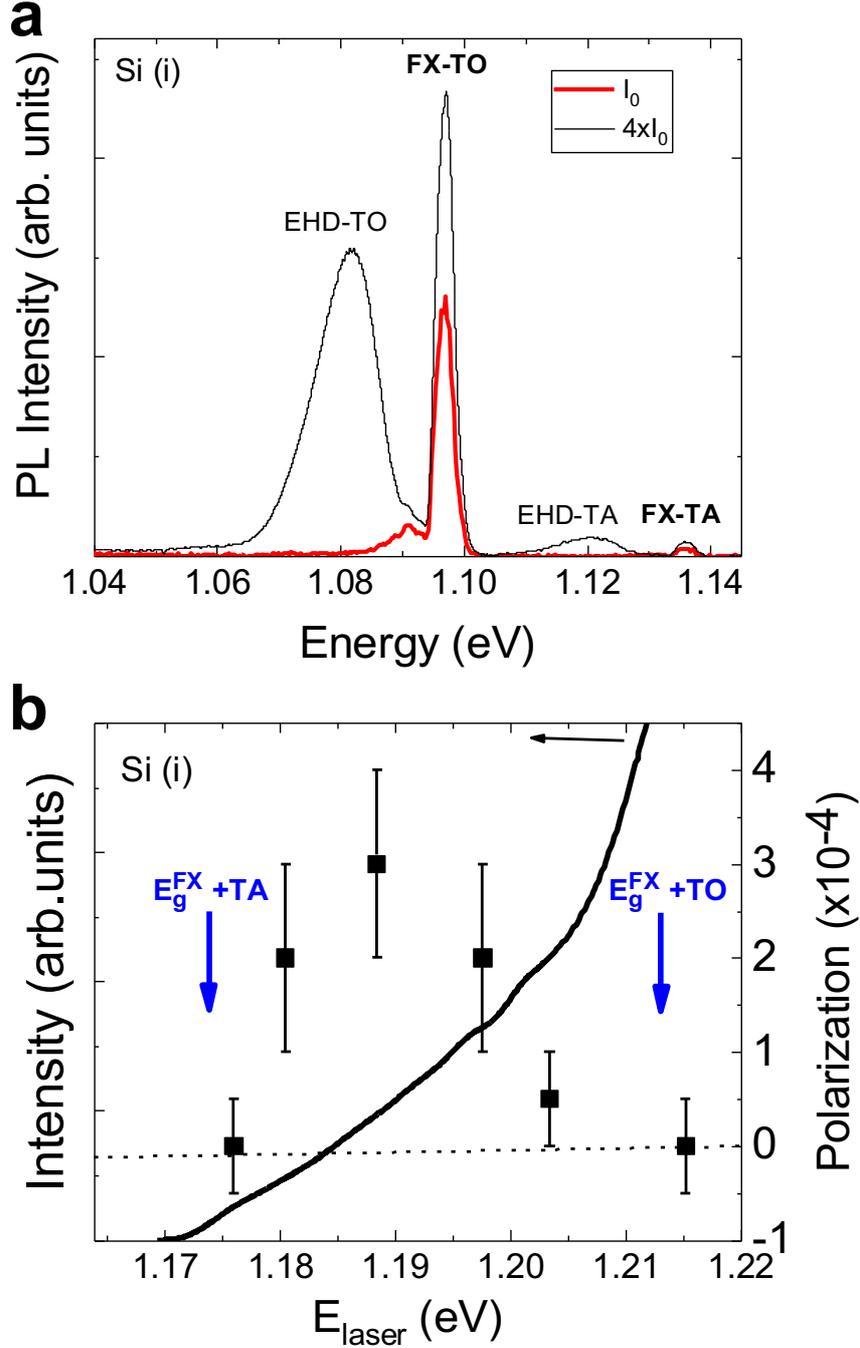

**Fig. S5 Optical spin orientation of <u>exciton</u>, indirect gap laser excitation - Intrinsic Silicon - (a)** Continuous wave photoluminescence spectra for two excitation intensities $I_0$= 5kW.cm$^{-2}$ and 4x$I_0$. The excitation laser energy is 1.96 eV. The narrow peak corresponds to the TO-assisted free exciton recombination line (FX-TO), whereas the broader line at lower energy detected for large excitation intensity results from the emission of electron-hole droplet (EHD-TO). TA-assisted emission of free exciton (FX-TA) and electron-hole droplet (EHD-TA) are also observed. **(b)** Variation of the luminescence polarization of the exciton luminescence (FX-TO) as a function of the excitation laser energy. The full line displays the variation of the FX-TO luminescence intensity (the blue vertical arrows indicate the onset of TO and TA assisted phonon exciton absorption : $E_g^{FX} + TO$ and $E_g^{FX} + TA$ [6].

**Section 6. Polarized photoluminescence in a large longitudinal magnetic field**

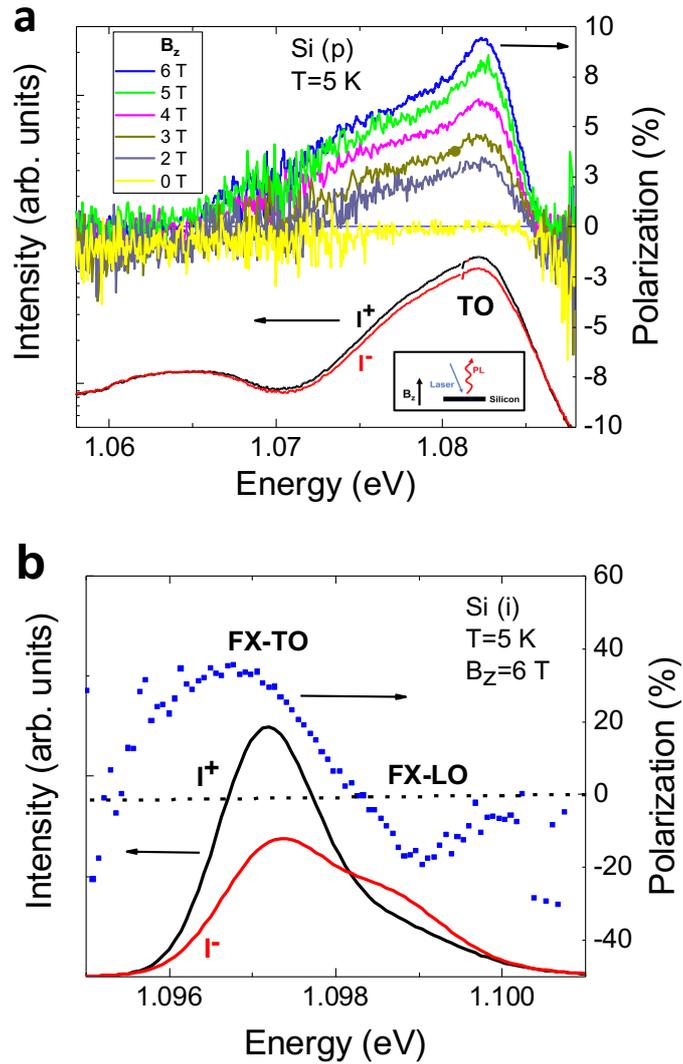

**Fig. S6. Polarized photoluminescence in a large longitudinal magnetic field**. The magnetic field $B_z$ is perpendicular to the sample plane and parallel to the light propagation axis, Faraday configuration (see the inset). In contrast to all the previous figures, the excitation laser (energy 1.96 eV) is linearly polarized and not circularly polarized. The measured circular polarization results here from the thermal carrier spin polarization induced by the Zeeman splitting and not from optical spin pumping [15] **(a) – p-doped Silicon –** The black and red lines are the right ($I^+$) and left ($I^-$) circularly-polarized luminescence intensity (measured for $B_z= 6$ T) ; the upper curves correspond to the measured circular polarization $P_c=(I^+-I^-)/(I^++I^-)$ of the TO phonon-assisted recombination, for magnetic fields $B_z$ varying between 0 and 6 T. No polarization is observed as expected for $B_z=0$ whereas significant polarization as large as 9% shows up for $B_z=6$ T, in agreement with previous measurements [16]. This demonstrates that the spin polarization effect related to the electron Zeeman splitting can play an important role for the measurement of circular polarization of silicon luminescence when a magnetic field of several Tesla is applied [17]. **(b) - Intrinsic Silicon -** The black and red lines are the right ($I^+$) and left ($I^-$) circularly-polarized luminescence intensity of the TO and LO-assisted recombination of free exciton (FX) ; the blue dots correspond to the measured circular polarization $P_c=(I^+-$

I$^-$)/(I$^+$+I$^-$) for $B_z$= 6 T. The negative polarization observed at 1.099 eV results directly from the optical selection rules associated to the LO-phonon assisted recombination (see Table I).